\documentstyle[twocolumn]{jpsj}
\def\runtitle{Physical Validity of Assumptions for Public Safety}
\def\runauthor{Tsuyoshi HONDOU}
\title{Physical Validity of Assumptions for Public Exposure to Mobile Phones}
\def\runauthor{Tsuyoshi HONDOU}

\def\runtitle{Validity of physical assumptions for public exposure to mobile phones}
\author{%
 Tsuyoshi \textsc{Hondou}\thanks{E-mail address: hondou@cmpt.phys.tohoku.ac.jp}}

\inst{% 
       Department of Physics, Tohoku University, Sendai 980-8578,  Japan
}

\kword{%
free energy, boundary condition, non-thermal interaction, mobile phone,
public safety, precautional principle, security hole}

\begin{document}
\maketitle

In the Letter\cite{Hondou}, we derived an analytical  formula which estimates the level of public exposure 
to electromagnetic waves in closed areas. This is the first study in which one approximately predicts
how much the exposure level increases by using two indispensable factors; 1) reflection of the electromagnetic
waves at the boundary and 2) additivity of emissions.
In their  Comment,  Kramer {\it et al.} claimed that  public exposure
in closed areas does not impose 
additional health risks in comparison with  those in any other location. 
Their claim was not toward the analytical derivation of the exposure level, which is the primary result of the Letter.
 However, we found that the Comment
was based on several implicit assumptions which were not relevant to the issue being addressed and thus
the claim itself was not valid. Some of the assumptions were based on improper applications of the  fundamentals of 
physics.  In this Reply,
we will clarify such misunderstandings  arising  in the Comment through careful consideration of their
implicit assumptions. The readers will find the essential issues which should always be taken into account.

First, we wish to reconfirm the absolutely essential requirements to deal with issues of public safety.
To  prevent {\it security holes} in  public safety, one is required to have a sufficiently deep insight into all the issues concerned,
 because a one-sided way
of looking at things may cause unexpected problems. 
 It  would be best 
 to start discussions from the fundamentals of physics which 
 relate to the issues.
A confusion between two distinct physical quantities can often be seen in the studies of
exposure, where electromagnetic energy is often equated with thermal energy.
As a fundamental fact, it is known that the energy of an electromagnetic wave is different from thermal
energy, as immediately proved by the second law of thermodynamics\cite{Callen}. Thermodynamics reveals that
{\em any} 
physical system can develop autonomously only in the direction in which the entropy of the system increases.
   One can fully change the energy of an electromagnetic 
wave into thermal energy, however the reverse is impossible\cite{note10}. It is a universal  fact
that 
the status of the energy of an electromagnetic wave is much more coherent than that of thermal energy. 
 The difference in energy can be 
described as, {\em free energy} or {\em entropy}\cite{note99}.
We can learn from this consideration that it is a natural phenomenon
 that electromagnetic waves often 
interfere with living things to a much large extent than does thermal energy under 
exposure to heat even if the apparent
quantities of energy are the same.

  It is  known that there are two kinds of electromagnetic interference for humans caused by electromagnetic waves,
1) thermal effect (heating) which is  attributed to an increase in temperature as a result of 
the dissipation process of electromagnetic waves in the human body,
and 2) non-thermal effect which is  attributed to the property of the electromagnetic wave itself.
The quantity Kramer {\it et al.} referred to in their Comment when they discussed safety issues
 was the specific absorption rate (SAR) 
 which 
concerns the thermal effect;  however, the non-thermal effect is indeed indispensable and must
be
considered, as stated in the last paragraph.  For example, electromagnetic waves may interfere with
the cardio-pacemakers implanted in cardiac patients. In the worst case of the interference,
 the person may pass away when such an interference
continues for several minutes. Note that a legal medical issue remains that no evidence would exist 
after the interference disappears.
The authors of the Comment completely ignored these serious issues, whereas the issue
 is explicitly stated in the Letter.
 The interference with pacemakers is reported to occur within a maximum distance of 30 cm between
a mobile phone and pacemaker\cite{JAPAN}, even though  neither reflectivity of the boundary nor
additivity of multiple phones has been taken into account in this experiment. 
In fact, the interference with pacemakers  
may occur even if the SAR limits (basic restriction by ICNIRP)  for environmental exposures,
an issue raised by the authors of the Comment, are cleared.
 This fact is also {\it explicitly stated}  in the section ``Purpose and Scope" of the document by ICNIRP 
that {\it Compliance with the present guidelines may not necessarily preclude interference with, or effects on,
medical devices $\cdots$. Interference with pacemakers may occur at levels below the recommended reference
levels. Advice on avoiding these problems is beyond the scope of the present document $\cdots$.} (see Ref. 2 
of the Comment). It is obvious that the interference with pacemakers is caused by 
non-thermal interference, the mechanism of which is physically different from 
the thermal effect regulated by SAR criteria. 
 The interference with hearing aids is also an example of current-day problems, where 
sufferers cannot use their hearing aids in certain places including commuter trains due to
insufferable noise caused by the electromagnetic interference.
Consideration of these examples reveals the limitation of the validity of the SAR criteria. SAR criteria 
are only applicable for recognized thermal effects and thus are not relevant 
to cover whole aspects of the safety issues by itself.

   The above discussion has already proved that the central claim of the Comment is not valid. 
 However, there remain crucial issues to be resolved for future studies. 
 The authors of the Comment introduced the present SAR criteria without any consideration of the
  limitation of validity.
What is more important for scientists is not the value of the criteria themselves but
their  scientific basis,
because such criteria can hardly be adopted without arbitrary factors. 
The criteria were derived through the review of extensive studies.
As stated in reference 7,
the basic restriction of ICNIRP (for microwave region) is based only on the study on the thermoregulatory response
caused by the thermal absorption of electromagnetic waves. There are
several other aspects regarding the origin of  the problem even within the thermal effects. 
It should also be noted that the exposure level is often  much lower than the well-recognized
thermal effect in case of non-thermal interaction, which includes not only the interference with
electromedical instruments but also the interference with living organisms themselves (biological effect). 
 SAR criteria of the basic restriction were, however, based only on clear evidence at the time 
of submission (see, ref. 7).
A number of studies, which had uncertainties in their results at the time, were not adopted 
in the basic restriction. Some of them reported adverse biological effects at a level
much lower than that adopted for the basic restriction.
However such not-well-recognized phenomena including non-thermal effects, were {\it not} considered
even as
``factors" to strengthen the basic restriction of ICNIRP, although
there is a gap between the extent of our knowledge and what is actually occurring. Namely, in a scientific sense,
the present SAR limit (basic restriction) does not guarantee that public safety is satisfied
under the regulation. 
 This background
justifies the  ``precautional principle" adopted in European countries where more strict regulations of the 
exposure level than that of ICNIRP have been adopted in order to compensate
for our ignorance on matters related to nature.

 Furthermore, there are more implicit assumptions made in the discussion in the Comment, which 
should have been explicit: 
(I) The estimation of the SAR value (25\%) was made based on the assumption of equal thermal 
absorption for all 
persons: They neglected the ``focus" phenomenon
that  the intensity of the wave can be concentrated due to the geometry of the reflective
boundary condition. An example of this phenomenon can be found in a concert hall, where the loudness of the sound 
is often concentrated at special locations (seats) due to the reflection of the sound.
This is a universal and essential physical property of  {\it waves}.
(II) Square distance decay of energy flux was assumed:  It is among the fundamental requirements
 that we have to specify a 
boundary condition to solve any problem of electromagnetism\cite{phillips}. Without a boundary condition, we 
cannot obtain a solution for electromagnetism. The authors seem 
to have implicitly assumed a free boundary condition
for these problems, as the boundary condition was not specified in their self-reference
 (Ref. 4) of the Comment.
 The boundary condition of systems considered in the Letter is, however, completely
 different from those 
in which a free boundary condition is appropriate.  It is explicitly emphasized
in the Letter that ``short-range interaction paradigm" is no longer appropriate.
(III) Mobile phones were assumed only to be used close to the head: 
 This assumption is not always appropriate.
People sometimes use mobile phones away from the head and close to stomach when they use an internet service
 such as {\it i-mode}; furthermore the users may include pregnant women.
 (IV) The number of mobile phones per person was assumed to be one:
    In several countries including Japan, many people possess multiple phones, for example, 
    one for personal use 
    and the other for business use.  We have to consider the fact
that the situation is changing and is no longer the conventionally assumed one. 
 (V) A similar condition of human health was implicitly assumed between a person
who uses a mobile phone and a person who suffers from interference:
    The user uses a mobile phone at his/her own risk for the sake of convenience.
 If he/she feels an adverse health condition,
he/she can stop using the phone. However, a person who is exposed to the interference 
cannot control the existing exposure by him/herself.
The person may be equipped with a pacemaker, for example. 
Exposure to general public must be considered separate from that to the users themselves.
 
 As discussed above, the Comment were based on naive implicit assumptions
which are neither relevant
nor valid and sometimes with {\it fatal} consequences. This leads to several
 security holes in the regulation of public safety.
Therefore the central claim of the Comment
should be rejected.
We hope that this correspondence between the Comment and the Reply can aid
the reader in understanding the present issues and can contribute to future studies.


\begin{thebibliography}{99}
\bibitem{note10}
In this paper, we assume that the frequencies of electromagnetic waves are within a certain
range. As found in black body radiation, thermal energy can easily be converted into electromagnetic energy.
However, electromagnetic waves emitted from the black body is qualitatively different from the electromagnetic waves considered in this paper.
Black body cannot emit coherent electromagnetic waves with special frequency. The difference can be described by 
 entropy.
\end{thebibliography}
\end{document}